\documentclass[11pt,a4paper,twoside,groupcitations]{article}
\usepackage[]{latexsym,amsmath,amssymb,upgreek}
\usepackage[]{epsfig}
\usepackage{color}
\usepackage{braket}
\usepackage{graphicx}
\usepackage{bm}
\usepackage{cite}
\newcommand{\be}{\begin{eqnarray}}
\newcommand{\ee}{\end{eqnarray}}
\def\beq{\begin{equation}}
\def\eeq{\end{equation}}

\newcommand{\pro}[2]{\mbox{$\langle\, #1 \mid #2\,\rangle$}}

\newcommand{\lp}{\ell_{\rm p}}
\newcommand{\mpl}{m_{\rm p}}
\newcommand{\gn}{G_{\rm N}}
\newcommand{\rh}{r_{\rm H}}
\newcommand{\Rh}{R_{\rm H}}
\newcommand{\dd}{\mbox{${\rm d}$}}

\title{\bf Hair and entropy for slowly rotating quantum black holes}
\author{Wenbin~Feng$^{ab}$\thanks{E-mail: wenbin.feng@studio.unibo.it},
$\ $
Roldao~da~Rocha$^{c}$\thanks{E-mail: roldao.rocha@ufabc.edu.br},
$\ $
and
Roberto~Casadio$^{ab}$\thanks{E-mail: casadio@bo.infn.it}
\\
\\
$^a${\em Dipartimento di Fisica e Astronomia, Universit\`a di Bologna}
\\
{\em via Irnerio~46, 40126 Bologna, Italy}
\\
\\
$^b${\em I.N.F.N., Sezione di Bologna, I.S.~FLAG}
\\
{\em viale B.~Pichat~6/2, 40127 Bologna, Italy}
\\
\\
$^c${\em Federal University of ABC, Center of Mathematics}
\\
{\em Santo Andr\'e, 09210-580, Brazil.}
}
\date{}
\begin{document}
\maketitle
\begin{abstract}
We study the quantum hair associated with coherent states describing slowly rotating
black holes and show how it can be naturally related with the Bekenstein-Hawking entropy
and with 1-loop quantum corrections of the metric for the (effectively) non-rotating case.
We also estimate corrections induced by such quantum hair to the temperature 
of the Hawking radiation through the tunnelling method.
\end{abstract}
\section{Introduction}
\label{S:intro}
\setcounter{equation}{0}
The breakthrough in gravitational wave astronomy from the LIGO and Virgo
collaboration~\cite{LIGOScientific:2016aoc} has opened up a new observational window,
allowing us to directly learn more about black holes.
As solutions to the Einstein equations, these spacetimes contain singularities which
might just signal the breakdown of classical physics in the strong field regime.
In recent years, several ways of describing quantum aspects of black holes have been
proposed in the literature.
Some approaches, like the corpuscular picture~\cite{DvaliGomez}, assume that the geometry
should only emerge at suitable (macroscopic) scales from the underlying (microscopic) quantum field
theory of gravitons~\cite{feynman,deser}.
Bekenstein's conjecture for the horizon area quantisation~\cite{bekenstein,Bekenstein:1995ju}
then naturally follows for the occupation number of gravitons is proportional to the
square of the ADM mass $M$~\cite{ADM} in units of the Planck mass $\mpl$.\footnote{We
shall often use units with $c=1$, $\gn=\lp/\mpl$ and $\hbar=\lp\,\mpl$,
where $\lp$ and $\mpl$ denote the Planck length and mass, respectively.}
\par
An improved description of nonuniform geometries can be obtained by employing coherent states of 
gravitons~\cite{Barnich:2010bu,Muck:2013use},
which then leads to necessary departures from the classical Schwarzschild
metric~\cite{Casadio:2021eio} (and thermodynamics~\cite{Casadio:2023pmh,Casadio:2022pla}).
In particular, the central singularity of the Schwarzschild black hole is replaced
by an integrable singularity~\cite{Lukash:2013ts} without Cauchy horizons.
The coherent state is built for a scalar field whose expectation value effectively describes
the geometry emerging from the (longitudinal or temporal) polarisation of the graviton
in the linearised theory.
A similar analysis for electrically charged spherically symmetric black holes was then 
shown to remove both the central singularity and the Cauchy horizon~\cite{Casadio:2022ndh}.
\par
The majority of black holes in nature are very likely to spin, which motivates investigating
quantum descriptions of black holes with non-vanishing specific angular momentum
$a=J/M$~\cite{Casadio:2023iqt}.
A complete description of axisymmetric Kerr black holes~\cite{Kerr:1963me} remains beyond our scope,
but this (conceptually and phenomenologically) important issue can be addressed 
for slow rotation by considering coherent states of gravitons similarly to the
spherically symmetric case.
In particular, we will focus on the quantum description of the approximate Kerr metric for
$|a|\ll \gn\,M$, which can be written as~\cite{Poisson:2009pwt}
\be
\dd s^2
\simeq
-\left(1 + 2\,V\right)
\dd t^2
+
\frac{\dd r^2}{1+ 2\,V}
-
\frac {4\,\gn\,M\,a}{r}\,\sin ^2 \theta\, \dd t\, \dd\phi
+
r^2\,\dd\Omega^2
\ ,
\label{metric}
\ee
where $\dd \Omega^2 =\dd \theta^2+ \sin^2\theta\,\dd\phi^2$.
In the above, the metric function
\be
V
=
V_M
+
W_a
\ ,
\label{VW}
\label{Newtonian gravitational potential with a correction}
\ee
where
\be
V_M
=
-\frac{\gn\,M}{r}
\label{Vs}
\ee
corresponds to the Schwarzschild metric~\cite{Schwarzschild:1916uq} for $a=0$,
and 
\be
W_a
=
\frac{a^2}{2\,r^2}
\ .
\label{Wa}
\ee
In the above stationary geometry, the possible event horizon is a sphere located at $r=\rh$
defined as the largest (real) solution of $1+2\,V=0$.
We shall find that the very existence of a quantum coherent state again requires
departures from the classical geometry (at least) near the (would-be) classical
central singularity.
This induces the presence of ``quantum hair'',~\footnote{The concept of quantum hair has been
explored through different quantum gravity frameworks, see e.g.~Refs.~\cite{Giddings:1993de}.}
which we will further connect with the Bekenstein-Hawking entropy~\cite{bekenstein},
the Hawking evaporation~\cite{Hawking:1975vcx}, 
and 1-loop quantum corrections to the metric obtained in the weak-field 
approximation (see~\cite{Frob:2021mpb} and references therein for earlier works).
Since the quantum corrected geometry obtained from coherent states is not perturbative
(in the ratio $\gn\,M/r$ or Planck constant), the latter result extends, and provides an independent
support for, perturbative calculations.
\par
In Section~\ref{S:coherent}, we first review the classical solutions of the Klein-Gordon equation
and show how coherent states of a massless scalar field on a reference flat spacetime 
associated to the vacuum can be used to reproduce a black hole geometry with small angular
momentum;
Section~\ref{S:hair} is devoted to studying the quantum hair of such coherent state black holes,
whose existence implies that information about the interior state is present outside the horizon;
the relation with the Bekenstein-Hawking entropy is derived in Section~\ref{S:entropy}, 
where corrections to the Hawking temperature are also estimated using the semiclassical
tunnelling methods; Section~\ref{S:conc} contains concluding remarks.
\section{Coherent quantum states for slowly rotating geometry}
\label{S:coherent}
\setcounter{equation}{0}
Like in Ref.~\cite{Casadio:2021eio}, the quantum vacuum is here assumed
to correspond to a spacetime devoid of matter and gravitational excitations.
Any classical metric should then emerge from a suitable (highly excited) quantum state.
A standard approach for recovering classical behaviours employs coherent states, 
which is generically motivated by their property of minimising the quantum uncertainty,
and is further supported by studies of electrodynamics~\cite{Muck:2013use,Muck:2013orm},
linearised gravity~\cite{Bose:2021ytn}, and the de~Sitter
spacetime~\cite{Dvali:2013eja,Giusti:2021shf}.
\par
In particular, we want to reproduce the slowly rotating stationary geometry~\eqref{metric}
as the full general relativistic extension of the Newtonian potential.
The latter can be derived from the longitudinal mode of gravitons in the linearised theory and,
like for the static case of Ref.~\cite{Casadio:2021eio}, we assume that this feature is preserved
in the stationary limit of full general relativity.
We therefore try and obtain the complete metric function~\eqref{VW} as the expectation value of
an effective free massless scalar field $\sqrt{\gn}\,\Phi=V$ satisfying the Klein-Gordon equation
\be
\Box \Phi
=
0
\ . 
\label{massless KG equation}
\ee
It is convenient to employ spherical coordinates in which a complete (normalised)
set of positive frequency solutions is given by
\be
u_{\omega\ell m}
=
\frac{e^{- i\, \omega\, t}}{\sqrt{2\,\omega}}\, j_{\ell}(\omega\,r)\, Y_{\ell m}(\theta, \varphi)
\ ,
\label{umodes}
\ee
where $j_{\ell}$ are spherical Bessel functions of the first kind, and
\be
Y_{\ell}^m
= 
(-1)^m
\sqrt{\frac{(2\,{\ell}+1)({\ell}-m)!}{4\,\pi\, ({\ell}+m)!}}
\,P_{\ell}^m(\cos\theta)\,e^{i\, m\, \varphi}
\ ,
\ee
are spherical harmonics of degree ${\ell}$ and order $m$,
$P_{\ell}^m$ being associated Legendre polynomials.
We recall that these solutions are orthonormal,~\footnote{See Appendix~\ref{A:conv}
for more details about the notation.}
\be
\left(u_{\omega\ell m }| u_{\omega'\ell' m'}\right)
=
\frac{\pi}{2\,\omega^2}\,
\delta(\omega-\omega')\, \delta_{\ell \ell'}\, \delta_{m m'}
\ ,
\quad
\left(u_{\omega\ell m }| u_{\omega'\ell' m'}^{\ast}\right)
=
0
\ ,
\label{(uu)}
\ee
in the Klein-Gordon scalar product
\be
\left(f_1| f_2\right)
=
i
\int \dd^3 x
\left(
f_1^{\ast}\, \partial_t  f_2
-f_2\,\partial_t f_1^{\ast} 
\right)
\ .
\ee
\par
The quantum theory is built by mapping the field $\Phi$ into an operator
expanded in terms of the normal modes (\ref{umodes}), 
\be
\hat\Phi
=
\sum_\ell \sum_{ m = - {\ell}}^{{ \ell} }
\frac{2}{\pi}
\int\limits_0^{\infty}
\omega^2
\,{\dd} \omega\,\sqrt{\hbar}
\left[
u_{\omega\ell m}\,
\hat{a}_{\ell m}(\omega)
+
u_{\omega\ell m }^{\ast}\,
\hat{a}^{\dagger}_{\ell m}(\omega) 
\right]
\ .
\label{the quantum field operator}
\ee
Likewise, its conjugate momentum reads
\be
\hat\Pi
=
i\,\sum_\ell \sum_{ m = - {\ell}}^{\ell}
\frac{2}{\pi}
\int\limits_0^{\infty}
\omega^3\,{\dd} \omega\,
\sqrt{\hbar}
\left[
u_{\omega\ell m}\,
\hat{a}_{\ell m}(\omega) 
-
u_{\omega\ell m }^{\ast}\,
\hat{a}^{\dagger}_{\ell m}(\omega) 
\right]
\ .
\label{the quantum field operator conjugate momentum}
\ee
These operators satisfy the equal-time commutation relations,
\be
\left[
\hat\Phi(t,r,\theta,\varphi),\hat\Pi(t,r',\theta',\varphi')
\right]
=
i\,\hbar\,
\frac{\delta(r-r')}{r^2}\,
\frac{\delta(\theta-\theta')}{\sin\theta}\,
\delta(\varphi-\varphi')
\ ,
\label{[PhiPi]}
\ee
provided the creation and annihilation operators obey the commutation rules
\be
\left[
\hat{a}_{\ell m}(\omega) , \hat{a}^{\dagger}_{\ell' m'}(\omega')
\right]
=
\frac {\pi} {2\,\omega^2}\,\delta(\omega-\omega')\,
\delta_{\ell \ell'}\, \delta_{m m'}
\ .
\label{[aa]}
\ee
The vacuum state is first defined by $\hat a_{\ell m}(\omega)\ket{0}=0$ for
all allowed values of $\omega$, $\ell$ and $m$, and a basis for the Fock space is 
constructed by the usual action of creation operators.
\subsection{Semiclassical metric function}
\label{SS:semi}
We seek a quantum state of $\Phi$ which effectively reproduces (as closely as possible)
the expected slow-rotation limit of the Kerr geometry~\eqref{metric}, that is
\be
\label{the classical potential}
\sqrt{\gn}
\bra{V}
\hat\Phi(t, r, \theta, \varphi) 
\ket{V}
\simeq
V(r)
\ .
\ee
We can build $\ket{V}$ as a superposition of coherent states satisfying
\be
\hat{a}_{\ell m}(\omega) 
\ket{g_{\ell m}(\omega)}
=
g_{\ell m}(\omega) \,e^{i \,\gamma_{\ell m}(\omega)} 
\ket{g_{\ell m}(\omega)}
\ ,
\label{coherent state}
\ee
where $g_{\ell m}=g_{\ell m}^*$ and $\gamma_{\ell m}=\gamma_{\ell m}^*$, so that
\be
\sqrt{\gn}
\bra{V}
\hat\Phi 
\ket{V}
&\!\!=\!\!&
\lp\,
\sum_\ell
\sum_{m=-\ell}^\ell
\frac{2}{\pi}\,
\int\limits_0^{\infty}
\omega^2\, \dd \omega\,
j_\ell(\omega\,r)\,
\frac{(-1)^m}{\sqrt{2\,\omega}}\,
\sqrt{\frac{(2\,{\ell}+1)({\ell}-m)!}{4\,\pi\, ({\ell}+m)!}}
\nonumber
\\
&&
\qquad\qquad\qquad\qquad
\times\, 
2\,\cos(\omega\,t-\gamma_{\ell m}+m\,\varphi)\,
P_{\ell}^m(\cos\theta)
\,g_{\ell m}(\omega)
\ .
\label{Volm}
\ee
Since the Kerr metric is stationary and axially symmetric, we impose that the phases
$\gamma_{\ell m}\simeq \omega\, t+m\,\varphi$.
Indeed, one could argue that recovering exact spacetime symmetries with such a limiting procedure
reflects the fact that no perfect isometries exist in nature~\cite{Casadio:2021eio}.
\par
The coefficients $g_{\ell m}$ can be determined by expanding the metric field $V$
on the spatial part of the normal modes~\eqref{umodes},
\be
V(r,\theta)
=
\sum_\ell
\sum_{m=-\ell}^\ell
\frac{2}{\pi}
\int\limits_0^{\infty}
\omega^2\, \dd \omega\,
j_{\ell}(\omega r)\,(-1)^m\,
\sqrt{\frac{(2\,\ell+1)\,(\ell-m)!}{4\,\pi\,(\ell+m)!}}\,
P_\ell^m(\cos\theta)
\,\tilde{V}_{\ell m}(\omega)
\ .
\label{expansion of field V}
\ee
By comparing the expansions~\eqref{Volm} and~\eqref{expansion of field V}, we obtain
\be
g_{\ell m}
=
\sqrt{\frac{\omega}{2}}\,
\frac{\tilde{V}_{\ell m}(\omega)}{\lp}
\ .
\label{value of V}
\ee
The coherent state finally reads
\be
\ket{V}
=
\prod_\ell
\prod_{m=-\ell}^\ell
e^{-N_{\ell m}/2}
\exp\left\{
\frac{2}{\pi}
\int\limits_0^{\infty}
\omega^2\, \dd \omega\,
g_{\ell m} (\omega) \, \hat a^{\dagger}_{\ell m} (\omega)
\right\}
\ket{0}
\ ,
\label{coherent state built in Fock space}
\ee
where
\be
N_{\ell m}
=
\frac{2}{\pi}
\int\limits_0^{\infty}
\omega^2\, \dd \omega
\left|g_{\ell m}(\omega )\right|^2
\ ,
\label{Nlm}
\ee
is the occupation number for the state $\ket{g_{\ell m}(\omega)}$.
We note in particular that $N_V=\sum_{\ell m} N_{\ell m}$ measures the ``distance'' of $\ket{V}$
from the vacuum $\ket{0}$ in the Fock space and should be finite~\cite{Casadio:2021eio}.
\subsection{Schwarzschild geometry}
\label{SS:schw}
For zero angular momentum, hence $a=W_a=0$, the metric function~\eqref{Vs} is obtained from 
\be
\tilde V_{00}
=
-
\frac{2\,\sqrt{\pi}}{\omega^2}\,\gn\,M
\ ,
\ee
so that the only contributions to the coherent state $\ket{V_M}$ are given by
the eigenvalues~\cite{Casadio:2021eio}
\be
g_{00}
=
-\sqrt{\frac{2\,\pi}{\omega^3}}\,
\frac{M}{\mpl}
\ ,
\label{g_00}
\ee
yielding the total occupation number 
\be
N_M
=
N_{00}
=
4\,\frac{M^2}{\mpl^2}
\int\limits_0^{\infty}
\frac{\dd\omega}{\omega}
\ .
\ee
The number $N_M$ diverges logarithmically both in the infrared (IR) and in the ultraviolet (UV).
In particular, the UV divergence arises from demanding a Schwarzschild geometry for all
values of $r>0$ and can be formally regularised by introducing a cut-off
$\omega_{\rm UV} \sim 1/R_{\rm s}$, where $R_{\rm s}$ can be interpreted as the finite radius
of a regular matter source~\cite{Casadio:2017cdv}.
\par
Note in fact that the static geometry we are reconstructing from the coherent state should
be completely determined by the energy-momentum of the matter source in general relativity,
like the Newtonian potential is fully determined by the energy density in the linearised theory.
The UV cut-off is therefore just a mathematically simple way of accounting for the fact that the very
existence of a proper quantum state $\ket{V_M}$ requires the coefficients $g_{00}=g_{00}(\omega)$
to depart from their purely classical expression~\eqref{g_00} for $\omega \to \infty$.
This departure from the classical expression would in turn be related with the actual state of matter in 
the interior of the black hole.
Since we aim at a general analysis of the geometry, we just demand that $R_{\rm s}\lesssim \Rh=2\,\gn\,M$
for a (quantum) black hole~\cite{Casadio:2021eio} and do not investigate the connection between the geometry
and possible matter sources any further here (see Refs.~\cite{Casadio:2020ueb,Casadio:2023ymt,Casadio:2023uqs}).
Likewise, we introduce a IR cut-off $\omega_{\rm IR}=1/R_{\infty}$ to account for the
necessarily finite lifetime $\tau \sim R_{\infty}$ of the system, and finally write
\be
N_M
=
4\,\frac{M^2}{\mpl^2}
\ln\!\left(\frac{R_{\infty}}{R_{\rm s}}
\right)
\ .
\label{Nm}
\ee
\par
The coherent state $\ket{V_M}$ so defined corresponds to a quantum-corrected metric function
\be
V_{{\rm q}M}
&\!\!\simeq\!\!&
\sqrt{\gn}
\bra{V_M}\hat\Phi\ket{V_M}
=
\frac{1}{\pi^{3/2}}
\int\limits_{\omega_{\rm IR}}^{\omega_{\rm UV}}
\omega^2\, \dd \omega\,
j_{0}(\omega r)\,
\tilde{V}_{00}(\omega)
\nonumber
\\
&\!\!\simeq\!\!&
-\frac{2\,\gn\,M}{\pi\,r}
\int_{R_{\infty}^{-1}}^{R_{\rm s}^{-1}}
\dd\omega\,
\frac{\sin(\omega\,r)}{\omega}
\nonumber
\\
&\!\!\simeq\!\!&
-\frac{\gn\,M}{r}
\left\{1 -
\left[ 1 - \frac {2} {\pi}\,{\rm Si}\!
\left(\frac {r} {R_{\rm s}}
\right)
\right]
\right\}
\ ,
\label{VqM}
\ee
where we let $\omega_{\text{IR}} = 1/R_{\infty} \to 0 $ and $ {\rm Si} $ denotes the sine
integral function.
The emerging quantum-corrected geometry is correspondingly given by~\footnote{For $R_{\rm s}\to 0$,
the term in square brackets in Eq.~\eqref{VqM} vanishes at any $r>0$ and the Schwarzschild
metric is formally recovered.}
\be
\dd s^2
\simeq
-
\left(1 + 2\,V_{{\rm q}M}\right)
\dd t^2
+
\frac{\dd r^2}{1 + 2\,V_{{\rm q}M}}
+
r^2\,\dd\Omega^2
\ ,
\label{metricVqM}
\ee
which was already analysed in Ref.~\cite{Casadio:2021eio}, where further details can be found.
\subsection{Slowly rotating black hole}
\label{SS:slow}
The classical metric~\eqref{metric} is characterised by an angular momentum of modulus
$\hbar\ll J=|a|\,M\ll \gn\,M^2$ oriented along the axis of symmetry, so that $J^z=J$ for $a>0$, and
by the metric function $W_a$ in Eq.~\eqref{Wa}.
We can now show that a quantum state that reproduces such a metric,
as closely as possible, like we discussed in the previous Section~\ref{SS:schw},
can be obtained by linearly combining the coherent state $\ket{V_M}$ of the
Schwarzschild geometry with a suitable coherent state $\ket{W_a}$.
\par
The normal modes~\eqref{umodes} are eigenfunctions of the angular momentum
operators $\hat{L}^2$ and $\hat{L}_z$ (in Minkowski spacetime) with eigenvalues $\hbar^2\,\ell\,( \ell + 1 )$
and $\hbar\,m$, respectively.
The expectation values of the angular momentum operators on the coherent state
$\ket{g_{\ell m}(\omega)}$ are therefore given by (see Appendix~\ref{A:Lg})
\be
J_{\ell m}
=
\bra{g_{\ell m}(\omega)} 
\sqrt{\hat{L}^2}
\ket{g_{\ell m}(\omega)}
=
\hbar\,\sqrt{\ell\left( \ell+1 \right)}
\left| g_{\ell m}(\omega)\right|^2
\ ,
\label{expansion of field J^2}
\ee
and
\be
J_{\ell m}^z
=
\bra{g_{\ell m}(\omega)} 
\hat{L}_z 
\ket{g_{\ell m}(\omega)}
=
\hbar\,m
\left| g_{\ell m}(\omega)\right|^2
\ .
\label{expansion of field J^z}
\ee
The total angular momentum for a superposition $\ket{W}$
of states $\ket{g_{\ell m}(\omega)}$ can be obtained as
\be
J
\equiv
\bra{W} \sqrt{\hat L^2}\ket{W}
=
\sum_{\ell>0}
\sum_{m=-\ell}^\ell
\frac{2}{\pi}
\int\limits_0^{\infty}
\omega^2\,\dd\omega\,
J_{\ell m}(\omega)
=
\sum_{\ell>0}
\hbar\,\sqrt{\ell\left( \ell+1 \right)}
\sum_{m=-\ell}^\ell
N_{\ell m}
\ .
\label{J/M}
\ee
Likewise,
\be
J^z
\equiv
\bra{W} \hat L_z\ket{W}
=
\sum_{\ell>0}
\sum_{m=-\ell}^\ell
\frac{2}{\pi}
\int\limits_0^{\infty}
\omega^2\,\dd\omega\,
J_{\ell m}^z(\omega)
=
\sum_{\ell>0}
\sum_{m=-\ell}^\ell
\hbar\,m\,N_{\ell m}
\ .
\label{Lz}
\ee
\par
Let us next consider coherent states defined by the eigenvalues 
\be
g_{\ell m}
=
C_{\ell m}\,\frac{\sqrt{2\,\pi}\,\lp^\alpha\,M}{\omega^{3/2-\alpha}\,\mpl}
\ ,
\label{glm}
\ee
where $C_{\ell m}$ are numerical coefficients that do not depend on $\omega$ and $\ell\ge 1$.
The corresponding occupation numbers~\eqref{Nlm} are given by
\be
N_{\ell m}
\simeq
\left\{
\begin{array}{ll}
C_{\ell m}^2\,N_M
&
{\rm for}\ \alpha=0
\\
\strut\displaystyle
4\,C_{\ell m}^2\,\frac{M^2}{\mpl^2}
\left[
\left(
\frac{\lp}{R_{\rm s}}
\right)^{2\,\alpha}
-
\left(
\frac{\lp}{R_{\infty}}
\right)^{2\,\alpha}
\right]
\qquad
&
{\rm for}\ \alpha\neq 0 
\ ,
\end{array}
\right.
\ee
where $N_M\sim M^2/\mpl^2$ is given in Eq.~\eqref{Nm}.
Note that the IR limit $R_{\infty}\to \infty$ is regular only for $\alpha>0$,
for which $N_{\ell m}\ll N_M$ if $R_{\rm s}\gg\lp$.
In this case, we can further approximate
\be
N_{\ell m}
\simeq
4\,C_{\ell m}^2\,\frac{M^2}{\mpl^2}
\left(
\frac{\lp}{R_{\rm s}}
\right)^{2\,\alpha}
\sim
C_{\ell m}^2
\ ,
\ee
where we considered $R_{\rm s}\sim \Rh$ for a black hole.~\footnote{All numerical factors can
be included in $C_{\ell m}$.}
Moreover, the modification~\eqref{expansion of field V} to the metric function is given by
\be
W_{\ell m}
&\!\!\simeq\!\!&
\lp\,
\frac{2}{\pi}\,
\int\limits_{\omega_{\rm IR}}^{\omega_{\rm UV}}
\omega^2\, \dd \omega\,
j_{\ell}(\omega\,r)\,
\frac{(-1)^m}{\sqrt{2\,\omega}}
\sqrt{\frac{(2\,\ell+1)\,(\ell-m)!}{\pi\,(\ell+m)!}}\,
P_{\ell}^{m}(\cos\theta)
\,g_{\ell m}(\omega)
\nonumber
\\
&\!\!\simeq\!\!&
\frac{\gn\,M}{r}
\left(\frac{\lp}{r}\right)^\alpha
\left[
C_{\ell m}\,
(-1)^m\,
\sqrt{\frac{(2\,\ell+1)\,(\ell-m)!}{(\ell+m)!}}\,
P_{\ell}^{m}(\cos\theta)\,
\frac{2}{\pi}
\int_{0}^{r/R_{\rm s}}
{z^{\alpha}}\,\dd z\,
j_{\ell}(z)
\right]
\ ,
\qquad
\label{Wlm}
\ee
where the integral in square brackets can be expressed in terms of regularised hypergeometric
functions [see Eq.~\eqref{JFa}].
We then see that the leading terms in the correction~\eqref{Wlm} are of the classical form
$W_a\sim r^{-2}$ in Eq.~\eqref{Wa} if $\alpha=1$.
\par
Finally, the contribution to the angular momentum satisfies the classicality conditions
\be
\hbar
\ll
J_{\ell m}
\simeq
\hbar\,\sqrt{\ell\,(\ell+1)}\,
N_{\ell m}
\simeq
\hbar\,m\,N_{\ell m}
\simeq
J^z_{\ell m}
\ ,
\label{hJzJ}
\ee
provided $m\simeq \ell$ and $\ell\,N_{\ell m}\sim \ell\,C_{\ell m}^2 \gg 1$.
We can build a coherent state $\ket{W_a}$ that reproduces the geometry~\eqref{metric}
by including different coherent states~\eqref{glm} with $\alpha=1$ and 
angular momentum numbers satisfying these conditions.
The rotation coefficient will then be given by
\be
\frac{\mpl}{M}
\ll
\frac{a}{\gn\,M}
\sim
\frac{\mpl^2}{M^2}\,
\sum_{\ell=1}^{\ell_{\rm c}}
\sqrt{\ell\left( \ell+1 \right)}\,N_{\ell \ell}
\sim
\frac{1}{N_M}\,
\sum_{\ell=1}^{\ell_{\rm c}}
\sqrt{\ell\left( \ell+1 \right)}\,N_{\ell \ell}
\lesssim
\delta_J
\ll
1
\ ,
\label{Lc}
\ee
where we introduced a parameter $\delta_J>0$ to define the slow rotation regime in terms
of a maximum value of $\ell$, denoted by $\ell_{\rm c}$.
\par
The inclusion of states $\ket{W}$ like the above will give rise to quantum-corrected geometries
\be
\dd s^2
\simeq
-\left(1 + 2\,V_{{\rm q}M}+2\,W_{{\rm q}a}\right)
\dd t^2
+
\frac{\dd r^2}{1+ 2\,V_{{\rm q}M}+2\,W_{{\rm q}a}}
-
\frac {4\,\gn\,J}{r}\,\sin ^2 \theta\, \dd t\, \dd\phi
+
r^2\,\dd\Omega^2
\ ,
\label{metricq}
\ee 
where $V_{{\rm q}M}$ is given in Eq.~\eqref{VqM}, $W_{{\rm q}a}\simeq W_{\ell m}$ 
in Eq.~\eqref{Wlm} and $J$ in Eq.~\eqref{J/M}.
These geometries do not entail a weak-field approximation but are perturbative in
the angular momentum contributions, that is in $J$ and $W_{{\rm q}a}$.
\section{Quantum hair}
\label{S:hair}
\setcounter{equation}{0}
Black hole solutions in general relativity are determined only by
the total mass, angular momentum, and electric charge (if present).
These uniqueness theorems~\cite{Heusler} strongly limit the information about the internal state
of a black hole that can be obtained by outside observers.
However, the situation changes when we consider the quantum description
of black holes given by coherent states already for the spherically
symmetric case of Section~\ref{SS:schw}.
In fact, the coherent states from which the geometry emerges as a mean field effect
cannot accommodate for perfect Schwarzschild spacetimes~\cite{Casadio:2021eio},
but they instead depend on the internal structure of the matter sources (classically)
hidden inside the horizon, as we recalled in Section~\ref{SS:schw}.
\par
The classical case of slow rotation was considered in Section~\ref{SS:slow},
where we assumed that the quantum states of the geometry only include specific coherent
states~\eqref{glm} with $\alpha=1$ satisfying the relations in Eq.~\eqref{hJzJ} for the angular momentum.
However, the possibility that other states contribute can only be limited 
from the condition of recovering the classical metric~\eqref{metric} within the experimental
bounds.
Their presence, on the other hand, will constitute a further example of
quantum hair~\cite{Giddings:1993de}, with departures from the classical geometry. 
\par
Instead of attempting a general analysis, we shall only consider states that 
violate one of the classicality conditions defined in Section~\ref{SS:slow} at a time.
In particular, we will study a)~quantum contributions with $J^z\simeq J$ but $\alpha>1$
inducing departures from $V_M$ smaller than $W_a$ at large $r$ in Section~\ref{SS:a>1}
and b)~modes with $\alpha=1$ and $a>0$ given by Eq.~\eqref{Lc} but such that $|J^z|\ll J$
in Section~\ref{SS:JzJ}.
\subsection{Quantum metric corrections}
\label{SS:a>1}
An explicit example of a coherent state which satisfies the classical conditions 
$J^z\simeq J$ for the angular momentum but leads to a geometry with terms
that fall off at $r\gg \Rh=2\,\gn\,M$ faster than $W_a$ in Eq.~\eqref{Wa} is built from
\be
g_{\bar\ell\bar\ell}
=
C_{\bar\ell}\,\frac{\sqrt{2\,\pi}\,\lp^\alpha\,M}{\omega^{3/2-\alpha}\,\mpl}
\ ,
\label{gll}
\ee
where $\alpha>1$ and $\bar\ell$ is a fixed integer value.
The hairy geometry can now be obtained from
\be
W_{\bar\ell\bar\ell}
\simeq
\frac{\lp}{\pi^2}\,
\int\limits_{\omega_{\rm IR}}^{\omega_{\rm UV}}
\omega^{3/2}\, \dd \omega\,
j_{\bar\ell}(\omega\,r)\,
\frac{2\,{\bar\ell}+1}{2^{\bar\ell-1/2}\,\bar\ell !}\,
(\sin\theta)^{\bar \ell}
\,g_{\bar\ell\bar\ell}(\omega)
\ ,
\label{hVolm}
\ee
where we used Eq.~\eqref{Pll}.
We thus find 
\be
W_{\bar\ell\bar\ell}
&\!\!\simeq\!\!&
C_{\bar\ell}\,\frac{\lp^\alpha\,\gn\,M}{\pi^{3/2}}\,
\frac{2\,{\bar\ell}+1}{2^{\bar\ell-1}\,\bar\ell !}\,(\sin\theta)^{\bar \ell}
\int\limits_{\omega_{\rm IR}}^{\omega_{\rm UV}}
{\omega^{\alpha}}\,\dd \omega\,
j_{\bar\ell}(\omega\,r)
\nonumber
\\
&\!\!\simeq\!\!&
C_{\bar\ell}\,\frac{\gn\,M}{\pi^{3/2}\,r}
\left(\frac{\lp}{r}\right)^\alpha
\frac{2\,{\bar\ell}+1}{2^{\bar\ell-1}\,\bar\ell !}\,(\sin\theta)^{\bar \ell}
\int_{0}^{r/R_{\rm s}}
{z^{\alpha}}\,\dd z\,
j_{\bar\ell}(z)
\nonumber
\\
&\!\!\sim\!\!&
\frac{\gn\,M}{r}
\left(\frac{\lp}{r}\right)^\alpha
\ ,
\label{WqM0}
\ee
where the integral is given in Eq.~\eqref{JFa}.
\par
We can in particular estimate the correction on the (unperturbed) Schwarzschild horizon at $r=\Rh$,
\be
W_{\bar\ell\bar\ell}(\Rh)
\sim
\left(\frac{\mpl}{M}\right)^\alpha
(\sin\theta)^{\bar \ell}
\ .
\label{WqaH}
\ee
Such corrections with different $\bar\ell$ represent purely axial perturbations on the horizon,
with vanishingly small amplitude for macroscopically large black holes of mass $M\gg \mpl$
provided $\alpha\gtrsim 1$.
\subsection{Quantum angular momentum}
\label{SS:JzJ}
States that lead to metric functions of the classical form $W_a$ in Eq.~\eqref{Wa} 
with $a$ given by Eq.~\eqref{Lc} but satisfy 
\be
J^z
\sim
\sum_{\ell=1}^{\ell_{\rm c}}
\sum_{m=-\ell}^\ell
m\,N_{\ell m}
\simeq
0
\label{aza}
\ee
can be simply obtained by assuming $|g_{\ell m}|=|g_{\ell -m}|$ so that $N_{\ell m}=N_{\ell -m}$.
As an example, we consider 
\be
g_{\bar\ell\bar\ell}
=
g_{\bar\ell-\bar\ell}
=
C_{\bar\ell}\,\sqrt{\frac{2\,\pi}{\omega}}\,\frac{M}{\mpl}
\ ,
\label{gl-l}
\ee
where $\bar\ell$ is again a fixed integer value and $\bar\ell\,C_{\bar\ell}^2$ is of the correct
size to yield a rotation parameter $a>0$ satisfying the bounds in Eq.~\eqref{Lc}. 
The metric correction is now given by
\be
W_{\bar\ell\bar\ell}
\simeq
\frac{\lp}{\pi^2}\,
\int\limits_{\omega_{\rm IR}}^{\omega_{\rm UV}}
\omega^2\, \dd \omega\,
j_{\bar\ell}(\omega\,r)\,
\frac{(-1)^{\bar\ell}+1}{\sqrt{2\,\omega}}\,
\frac{2\,{\bar\ell}+1}{2^{\bar\ell-1}\,\bar\ell !}\,
(\sin\theta)^{\bar \ell}
\,g_{\bar\ell\bar\ell}(\omega)
\ ,
\label{hVolm}
\ee
where we used the known relation~\eqref{Pl-l}.
For $\bar\ell$ odd the above expression vanishes, whereas for $\bar\ell$ even we
find twice the value in Eq.~\eqref{WqM0} with $\alpha=1$, that is  
\be
W_{\bar\ell\bar\ell}
\sim
\frac{\lp\,\gn\,M}{r^2}
\ .
\ee
\par
From the above few examples, it should be clear that one can engineer many different axially
symmetric configurations, all of which differ from the (slowly-rotating)
Kerr geometry only by terms of order $(\lp/r)^\alpha$ for $\alpha\ge 1$.
Of course, this ambiguity would be removed by computing the coherent state generated
by a given matter source, which is however supposedly hidden behind the horizon.
Moreover, we remark that such terms would result in a (slight) shift in the position $\rh$
of the event horizon with respect to the classical value $\Rh$.
\section{Entropy and evaporation}
\label{S:entropy}
\setcounter{equation}{0}
In the previous Sections, for simplicity, we have modelled the dependence of the
geometry from the internal structure of the black hole
by introducing cut-offs $\omega_{\rm IR}\sim 1/R_{\infty}$
and $\omega_{\rm UV}\sim 1/R_{\rm s}$ in momentum space and allowing
for contributions of angular momentum that have no classical counterpart.
Were we able to test the gravitational field with sufficient accuracy, for instance from the motion of
test particles and light in the outer region to the horizon, we could remove these uncertainties
and gather information about the interior of the black hole.
\subsection{Bekenstein-Hawking entropy}
\label{SS:entropy}
A common way to measure our ignorance about the actual state of a system is provided
by the thermodynamic entropy, which is obtained by counting the possible microstates
corresponding to a given macroscopic configuration.
For a Schwarzschild black hole, the Bekenstein-Hawking entropy~\cite{bekenstein}
\be
S_{\rm BH}
=
\frac{\pi\,\Rh^2}{\lp^2}
=
\frac {4\,\pi\,M^2} {\mpl^2}
\label{Sbh}
\ee
can be obtained~\cite{Casadio:2023pmh} by supplementing a pure coherent state of the
Schwarzschild geometry~\eqref{g_00} with the Planckian distribution of Hawking quanta
at the temperature~\cite{Hawking:1975vcx}
\be
T_{\rm H}
=
\frac{\mpl^2}{8\,\pi\,M}
\ .
\label{Th}
\ee
\par
Given a black hole of mass $M$, instead of one pure coherent state, we could consider all possible
states giving rise to (practically) indistinguishable semiclassical geometries with the mass $M$.
We can employ the total occupation number~\eqref{Nm}~\footnote{We just mention that the
same quantisation law is obtained for the ground state of a dust ball~\cite{Casadio:2023ymt}.}
of the corresponding coherent state to estimate the total number of microstates available
to build such configurations
as
\be
\mathcal N_{M}
\sim
\sum_{n=0}^{N_M}
\left(
\!\!
\begin{array}{c}
N_M
\\
n
\end{array}
\!\!
\right)
=
\sum_{n=0}^{N_M}
\frac{N_M !}{(N_M-n)!\,n!}
=
2^{N_M}
\ . 
\ee
The thermodynamic entropy is thus
\be
S_M
\propto
\ln(\mathcal N_M)
\sim
\left(\frac{M}{\mpl}\right)^2
\ ,
\label{SbhM}
\ee
which is clearly proportional to the Bekenstein-Hawking entropy~\eqref{Sbh}.
One can therefore envisage that the coherent states giving rise to Schwarzschild black 
hole geometries contain the precursors (or proxies) of the Hawking particles,
like in the original corpuscular picture~\cite{DvaliGomez}.
\subsection{Entropy and angular momentum}
\label{SS:entropyM}
We can also estimate the number of quantum states with angular momentum 
corresponding to geometric configurations that cannot be observationally distinguished 
from a non-rotating Schwarz\-schild black hole.
For this purpose, we can consider again a maximum angular momentum parameter
$\delta_J\ll 1$ such that configurations with
\be
\frac {J} {M\,\Rh}
\simeq
\frac{a}{\gn\,M}
\lesssim
\delta_J
\ ,
\label{constraintJ}
\ee
cannot be distinguished from the coherent state reproducing the quantum-corrected
Schwarzschild geometry~\eqref{VqM}.
Furthermore, we shall include in this count only those contributions of the form
in Eq.~\eqref{glm}, 
\be
g_{\ell m}
\sim
C_{\ell m}\,\frac{\omega^\alpha\,\lp^\alpha\,M}{\omega^{3/2}\,\mpl}
\ ,
\label{gl0a}
\ee
that violate both of the classicality conditions considered in Sections~\ref{SS:a>1} and~\ref{SS:JzJ},
that is $\alpha\gtrsim1$ and $0\le |m|\ll\ell$. 
\par
In particular, the contribution of modes with $m\simeq 0$ to the angular momentum~\eqref{J/M}
is approximately given by
\be
\frac{a_{\ell}}{\gn\,M}
\simeq
\ell\,\frac{\mpl^2}{M^2}\,N_{\ell 0}
\sim
\ell
\left(\frac{\lp}{R_{\rm s}}\right)^{2\,\alpha}
\ ,
\ee
in which we assumed $N_{\ell 0}\sim C^2_{\ell 0}\sim 1$ and $\ell\gg 1$.
Imposing the constraint~\eqref{constraintJ} on the total angular momentum,
\be
\sum_{\ell=1}^{\ell_{\rm c}}
\frac{a_\ell}{\gn\,M}
\sim
\ell_{\rm c}^2
\left(\frac{\lp}{R_{\rm s}}\right)^{2\,\alpha}
\lesssim
\delta_J
\ ,
\ee
yields
\be
\ell_{\rm c}
\lesssim
\left(\frac{R_{\rm s}}{\lp}\right)^{\alpha}
\sqrt{\delta_J}
\sim
\left(\frac{M}{\mpl}\right)^{\alpha}
\sqrt{\delta_J}
\ ,
\label{dJ}
\ee
where we again set $R_{\rm s}\sim\Rh$ for a black hole.
Upon allowing for the inclusion of modes $\ket{g_{\ell 0}}$ with $1\le \ell\le\ell_{\rm c}$,
we can estimate the degeneracy of the quantum black hole
given by the total number of possible combinations in angular momentum, that is
\be
\mathcal N_{\rm c}
=
\sum_{\ell=0}^{\ell_{\rm c}}
\left(
\!\!
\begin{array}{c}
\ell_{\rm c}
\\
\ell
\end{array}
\!\!
\right)
=
\sum_{\ell=0}^{\ell_{\rm c}}
\frac{\ell_{\rm c} !}{(\ell_{\rm c}-\ell)!\,\ell !}
=
2^{\ell_{\rm c}}
\ . 
\ee
The corresponding thermodynamic entropy,
\be
S
\propto
\ln(\mathcal N_{\rm c})
\sim
\left(\frac{M}{\mpl}\right)^{\alpha}
\sqrt{\delta_J}
\ ,
\label{Sg}
\ee
is also proportional to the Bekenstein-Hawking entropy~\eqref{Sbh} for $\alpha=2$.
\par
It is then interesting to notice that the metric corrections for $\alpha=2$ are
of the same order in $\gn$, $\lp$ and $1/r$ as those obtained from 1-loop corrections
to the Schwarzschild metric~\cite{Frob:2021mpb}, that is
\be
W_{{\rm q}a}
&\!\!\simeq\!\!&
\sum_{\ell=1}^{\ell_{\rm c}}\,W_{\ell 0}
\simeq
\frac{\gn\,M}{r}
\left(\frac{\lp}{r}\right)^2
\sum_{\ell=1}^{\ell_{\rm c}}
\left[
C_{\ell 0}\,
\sqrt{2\,\ell+1}\,
P_{\ell}(\cos\theta)\,
\frac{2}{\pi}
\int_{0}^{r/R_{\rm s}}
{z^{2}}\,\dd z\,
j_{\ell}(z)
\right]
\nonumber
\\
&\!\!\sim\!\!&
\frac{\gn\,M}{r}
\left(
\frac{\lp}{r}
\right)^2
\ ,
\label{Wl02}
\ee
where $P_\ell=P_\ell^0$ are Legendre polynomials.
The corresponding quantum corrected Schwarzschild geometry is now given by
\be
\dd s^2
\simeq
-
\left(1 + 2\,V_{{\rm q}M}+2\,W_{{\rm q}a}\right)
\dd t^2
+
\frac{\dd r^2}{1 + 2\,V_{{\rm q}M}+2\,W_{{\rm q}a}}
+
r^2\,\dd\Omega^2
\ ,
\label{metricVqMa}
\ee
where $V_{{\rm q}M}$ is the metric function in Eq.~\eqref{VqM}.
It is important to remark that $W_{{\rm q}a}$ represents a perturbation over the full
quantum-corrected geometry~\eqref{metricVqM} and is not restricted to the weak-field approximation
employed to perform 1-loop corrections.
\par
Finally, we can check that the condition~\eqref{dJ} guarantees that the horizon does not shift
significantly from the unperturbed Schwarzschild radius.
In fact, for $\alpha=2$ and neglecting the effect of $V_{{\rm q}M}$, we can write the metric 
function
\be
V
\simeq
V_M
+
\epsilon\,\frac{\lp^2\,\gn\,M}{r^3}
\ ,
\label{VMe}
\ee
where $\epsilon\sim\sqrt{\delta_J}$ now contains all the parameters (and angular dependence)
shown in the first line of Eq.~\eqref{Wl02}.  
The largest solution to $V=-1/2$ is then given by
\be
\rh
\simeq
2\,\gn\,M
-
\epsilon\,\lp
\ ,
\ee
which represents a negligible correction to $\Rh=2\,\gn\,M$.
Given the fast fall-off of the metric correction in Eq.~\eqref{Wl02}, one could interpret these perturbations
as being ``confined'' about the horizon $\Rh$, like in the membrane approach~\cite{Thorne:1986iy} and
in derivation of the entropy~\eqref{Sbh} based on conformal symmetry~\cite{Carlip:1998wz}.
\subsection{Hawking radiation}
\label{SS:Hrad}
The Hawking evaporation has been studied with several methods since its discovery~\cite{Hawking:1975vcx}.
In particular, semiclassical approaches describe this effect
as particles that tunnel out from within the event horizon on classically forbidden
paths~\cite{Parikh:1999mf,tunnel,Angheben:2005rm,Casadio:2017sze}.
We will employ the WKB approach to compute corrections to the Hawking temperature
for slowly rotating black holes described by the quantum-corrected Schwarzschild
metric~\eqref{VqM} with the metric modification in Eq.~\eqref{Wl02} that we showed
can contribute to the Bekenstein-Hawking entropy.
\par
We start by noting that, replacing the WKB ansatz
\be
\Phi
\simeq
\exp\left[\frac{i}{\hbar }\,S(t,r, \theta,\phi)\right]
\ee
in the Klein-Gordon Eq.~\eqref{massless KG equation} at leading order in $\hbar$,
yields the Hamilton-Jacobi equation
\be
g^{\mu \nu }\,\partial _{\mu }S\,\partial _{\nu }S
\simeq
0
\ .
\label{hj}
\ee
Solutions can be written in the form 
\be
S
=
-E\,t
+
\mathcal W(r)
+
\mathcal J(\theta,\phi)
+
K
\ ,
\label{Swjk}
\ee
where $E$ represents the energy of the emitted boson and $K$ is a complex constant that
will be fixed later.
The ratio $E/M$ regulates the magnitude of the backreaction of the emission on the black hole,
which can alter the thermal nature of the Hawking radiation~\cite{Parikh:1999mf}.
We only consider large black holes with mass $M\gg\mpl$, hence this effect can be
neglected for $E\ll M$.
\par
Given the inverse of the metric~\eqref{metric}, Eq.~(\ref{hj}) can be written as 
\be
{(1+2\,V)}
\left[
\left(\frac{\partial \mathcal W}{\partial r}\right)^2
+
\frac{r^2}{w\,\sin\theta}
\left(\frac{\partial \mathcal J}{\partial \phi}\right)^2
\right]
+
\frac{1}{r^2}\left(\frac{\partial \mathcal J}{\partial \theta}\right)^2
+
\frac{4 \,a \,\gn\,M\,r}{w} 
E\,\frac{\partial \mathcal J}{\partial \phi}
\simeq
\frac{r^4}{w}
E^2
\ ,
\label{solv}
\ee
where we used the form~\eqref{Swjk} for $S$ and defined
\be
w
\equiv
r^4\,(1+2\,V)
+4\, a^2\,\gn^2\, M^2\, \sin^2\theta
\ .
\label{w}
\ee
For Hawking particles in a quantum corrected Schwarzschild geometry,
we can just consider purely radial trajectories~\cite{Angheben:2005rm},
along which $\mathcal J$ is constant, and further approximate $w\simeq r^4\left(1+2\,V\right)$.
In this case, Eq.~\eqref{solv} is solved by
\be
\mathcal W_{\pm }
\simeq
\pm
E
\int^r
\frac{\dd \bar r}
{1+2\,V}
\ ,
\label{10}
\ee
with $+$ ($-$) for outgoing (ingoing) particles.
\par
Imaginary terms in the action $S$ correspond to the Boltzmann factor for emission
and absorption across the event horizon.
Such terms can only arise due to the pole at $r=\rh\simeq \Rh$, where $1+2\,V=0$,
and from the imaginary part of $K$ in Eq.~\eqref{Swjk}, resulting in the probabilities
\be
P_\pm
\propto
\exp\left[-\frac{2}{\hbar }\left(\Im \mathcal W_\pm+\Im K\right)\right]
\ ,
\label{outprob}
\ee
where $\Im$ denotes the imaginary part.
Assuming that ingoing particles necessarily cross the event horizon, that is $P_-\simeq 1$,
one must set $\Im K=-\Im \mathcal W_{-}$.
Since $\mathcal W_{+}=-\mathcal W_{-}$, the probability of a particle tunnelling out 
then reads 
\be
P_+
\simeq
\exp\left(-\frac{4}{\hbar }\,\Im \mathcal W_{+}\right)
\ .
\label{finprob}
\ee
The integral~\eqref{10} around the pole at $r\simeq\Rh$ with the Feynman prescription
for the propagator~\cite{Angheben:2005rm} yields
\be
\Im
\mathcal W_+
\simeq
\lim_{r\to \Rh}
\frac{\pi\, E}{2\,\hbar\, V'(r)}
\ ,
\label{FormW}
\ee
where $V'=\partial_r V$.
Finally, 
\be
P_+
\simeq
\exp\left[-\frac{2\,\pi\, E}{V'(\Rh)}\right]
\ 
\ee
which implies that the temperature must be given by
\be
T_M
\simeq
\frac{\hbar}{2\,\pi }\,
V'(\Rh)
=
\frac{\hbar}{2\,\pi }
\left[
V_{{\rm q}M}'(\Rh)
+
W_{{\rm q}a}'(\Rh)
\right]
\ .
\label{Hawk}
\ee
This expression with the metric function~\eqref{VqM} and the contribution~\eqref{Wl02} with $\alpha=2$ for the case of
Section~\ref{SS:entropy} gives
\be
T_M
\simeq
T_{\rm H}
\,\frac{2}{\pi}
\left[
{\rm Si}\left(\frac{R_{\rm H}}{R_s}\right)
-
\sin\!\left(\frac{\Rh}{R_s}\right)
-
\frac{3\,\mpl^2}{4\,\pi\,M^2}\,
\sum_{\ell=1}^{\ell_c} 
C_{\ell0}\,\sqrt{2\,\ell+1}\,
P_\ell(\cos\theta)
\!\!\!
\int\limits_0^{R_{\rm H}/R_s} 
\!\!\!
z^2\,\dd z\,j_\ell(z)
\right]
\ ,
\label{13}
\ee
where $T_{\rm H}$ is the standard Hawking temperature~\eqref{Th},
which is therefore recovered asymptotically for $M\gg \mpl$ and $R_{\rm s}\ll \Rh$.
\par
Using the metric function in Eq.~\eqref{VMe}, one analogously finds
\be
T_M 
\simeq
T_{\rm H}
\left(
1
- \frac{3\,\epsilon\,\mpl^2}{4\,M^2}
\right)
\ .
\label{16}
\ee
On equating the two corrections of order $\mpl^2/M^2$, we obtain
\be
\epsilon
\simeq
\frac{1}{\sqrt{\pi^3}}\,
\sum_{\ell=1}^{\ell_c} 
C_{\ell0}\,
\frac{\sqrt{2\,\ell+1}\,\Gamma(\ell/2+3/2)\,P_\ell(\cos\theta)}
{{2^{\ell}}\,\Gamma\left(\ell+3/2\right)\,\Gamma(\ell/2 +5/2)}\,
{}_1F_2\left(\frac{\ell+3}{2},\ell+\frac{3}{2},\frac{\ell+5}{2}, -\frac{1}{4}\right)
\ ,
\ee
where we used Eq.~\eqref{JFa} with $\alpha=2$.
\section{Conclusions and outlook}
\label{S:conc}
\setcounter{equation}{0}
In this work, the semiclassical metric function reproducing a Kerr geometry in the slow-rotation
regime was shown to arise from suitable highly-excited coherent states, thus generalising previous results
obtained for spherically symmetric geometries~\cite{Casadio:2021eio,Casadio:2022ndh,Giusti:2021shf}.
Quantum hair naturally emerges in this context, since the existence of the quantum coherent state
does not allow for any possible IR and UV divergences in general.
\par
An additional source of quantum hair was then identified in angular momentum modes that do not satisfy
the conditions for giving rise to a classical rotating geometry described in Section~\ref{SS:slow}.
Such modes were further associated with the Bekenstein-Hawking entropy of Schwarzschild black holes
and are therefore expected to play the role of precursors of the Hawking radiation, at least
for very massive black holes.
The Hawking evaporation was then studied with the Hamilton-Jacobi method, from which
modes representing quantum hair in the geometry were related to metric corrections of the
form that one expects from 1-loop quantum corrections in the weak-field 
approximation~\cite{Frob:2021mpb}.  
\par
There are different directions along which the present results could be improved and
developed.
First of all, results regarding the Hawking evaporation can be straightforwardly generalised
to massive bosons and fermions~\cite{Casadio:2017sze}.
One could furthermore study other black hole solutions that can emerge from coherent
quantum states and eventually attempt at a quantum description of black holes with arbitrary
angular momentum~\cite{Casadio:2023iqt}.
\section*{Acknowledgments}
W.F.~acknowledges the financial support provided by the scholarship granted by the
Chinese Scholarship Council (CSC).
W.F.~and R.C.~are partially supported by the INFN grant FLAG.
The work of R.C.~has also been carried out in the framework of activities of the
National Group of Mathematical Physics (GNFM, INdAM).
R.dR.~is grateful to FAPESP (Grants No.~2022/01734-7 and No.~2021/01089-1),
the National Council for Scientific and Technological Development--CNPq
(Grant No.~303390/2019-0), and the Coordination for the Improvement of Higher Education
Personnel (CAPES-PrInt~88887.897177/2023-00) for partial financial support.
R.dR.~thanks R.C.~and DIFA, Universit\`a di Bologna, for the hospitality. 
\appendix
\section{Normalisations and conventions}
\label{A:conv}
\setcounter{equation}{0}
We summarise here the convention we use in the main text.
Projections on the spatial part of the normal modes~\eqref{umodes} are defined as
\be
\tilde f_{\ell m}(\omega)
=
\int\limits_{-1}^{+1}
\dd\cos\theta
\int\limits_0^{2\,\pi}
\dd\varphi
\int\limits_0^\infty
r^2\,\dd r\,
j_\ell(\omega\,r)
\left[Y_{\ell}^m(\theta,\varphi)\right]^*
f(r,\theta,\varphi)
\ .
\ee
The orthonormality relations~\eqref{(uu)} then follow
from the orthonormality of spherical Bessel functions,
\be
\int\limits_0^\infty
r^2\,\dd r\,j_\ell(\omega\,r)\,j_{\ell'}(\omega'\,r)
=
\frac{\pi}{2\,\omega^2}\,
\delta(\omega-\omega')
\,\delta_{\ell\ell'}
\ ,
\ee
as well as the orthonormality of spherical harmonics,
\be
\int\limits_{-1}^{+1}
\dd\cos\theta
\int\limits_0^{2\,\pi}
\dd\varphi\,
Y_\ell^m(\theta,\varphi)
\left[Y_{\ell'}^{m'}(\theta,\varphi)\right]^*
=
\delta_{\ell\ell'}\,\delta_{m m'}
\ .
\ee
The commutation relations~\eqref{[PhiPi]} and~\eqref{[aa]}
follow from the completeness relations
\be
\frac{2}{\pi}
\int\limits_0^\infty
\omega^2\,\dd\omega\,
j_\ell(\omega\,r)\,j_{\ell}(\omega\,r')
=
\frac{\delta(r-r')}{r^2}
\ee
and
\be
\sum_\ell
\sum_{m=-\ell}^\ell
Y_\ell^m(\theta,\varphi)
\left[Y_\ell^m(\theta',\varphi')
\right]^*
=
\frac{\delta(\theta-\theta')}{\sin\theta}\,
\delta(\varphi-\varphi')
\ .
\ee
Other useful properties of spherical harmonics are given by
\be
\left[Y_\ell^m\right]^*
=
(-1)^m\,Y_\ell^{-m}
\label{Y-Y}
\ee
and
\be
P_\ell^{-m}=(-1)^m\,\frac{(\ell-m)!}{(\ell+m)!}\,P_\ell^{m}
\ .
\ee
From
\be
P_\ell^\ell
=
\frac{(-1)^\ell}{2^\ell\,\ell !}
\sqrt{\frac{(2\,\ell+1)!}{4\,\pi}}\,
(\sin\theta)^\ell
\ ,
\label{Pll}
\ee
we then obtain
\be
P_\ell^{-\ell}
=
\frac{1}{2^\ell\,\ell!\,(2\,\ell)!}
\sqrt{\frac{(2\,\ell+1)!}{4\,\pi}}\,
(\sin\theta)^\ell
\ .
\label{Pl-l}
\ee
\par
In all of the above expressions, the Kronecker delta is defined by $\delta_{ij}=1$ for $i=j$ 
and $\delta_{ij}=0$ for $i\neq j$.
The Dirac delta is defined by
\be
\int
\dd z\,\delta(z-z_0)\,f(z)
=
f(z_0)
\ ,
\ee
where integration is assumed on the natural domain of the
variable $z$.
\par
Relevant integrals of the spherical Bessel functions are given by
\be
\int_0^x
z^\alpha\,\dd z\,j_\ell(z)
&\!\!=\!\!&
\frac{\sqrt{\pi}}{2^{\ell+2}}\,
\frac{\Gamma((1+\alpha+\ell)/2)}{\Gamma(3/2+\ell)\,\Gamma((3+\alpha+\ell)/2)}
\nonumber
\\
&&
\times
{_1 F_2}((1+\alpha+\ell)/2,\ell+3/2,(3+\alpha+\ell)/2,-x^2/4)
\ ,
\label{JFa}
\ee
where ${_1 F_2}$ is the generalised hypergeometric function.
In particular, for $\alpha=1$, we have
\be
\int_0^x
z\,\dd z\,j_\ell(z)
=
\frac{\sqrt{\pi}}{2^{\ell+2}}\,
\frac{\Gamma(1+\ell/2)}{\Gamma(3/2+\ell)\,\Gamma(2+\ell/2)}\,
{_1 F_2}(1+\ell/2,\ell+3/2,2+\ell/2,-x^2/4)
\ .
\label{JFa1}
\ee
\section{Angular momentum}
\label{A:Lg}
\setcounter{equation}{0}
The normal modes~\eqref{umodes} are eigenfunctions of the angular momentum,
that is
\be
\hat L^2\,
u_{\omega\ell m}
=
\hbar^2\,\ell\left(\ell+1\right)
u_{\omega\ell m}
\qquad
{\rm and}
\qquad
\hat L_z\,
u_{\omega\ell m}
=
\hbar\,m\,
u_{\omega\ell m}
\ .
\ee
It then follows that
\be
\hat L^2
\ket{1_{\ell m}(\omega)}
=
\hbar^2\,\ell\left(\ell+1\right)
\ket{1_{\ell m}(\omega)}
\qquad
{\rm and}
\qquad
\hat L_z
\ket{1_{\ell m}(\omega)}
=
\hbar\,m
\ket{1_{\ell m}(\omega)}
\ ,
\ee
where $\ket{1_{\ell m}(\omega)}=\hat a^\dagger_{\ell m}(\omega)\ket{0}$.
We can also write the first relation as defining the operator
\be
\sqrt{\hat L^2}
\ket{1_{\ell m}(\omega)}
=
\hbar\,\sqrt{\ell\left(\ell+1\right)}
\ket{1_{\ell m}(\omega)}
\ .
\ee
Likewise, we have
\be
\sqrt{\hat L^2}
\ket{n_{\ell m}(\omega)}
=
\hbar\,\sqrt{\ell\left(\ell+1\right)}\,
n_{\ell m}
\ket{n_{\ell m}(\omega)}
\quad
{\rm and}
\quad
\hat L_z
\ket{n_{\ell m}(\omega)}
=
\hbar\,m\,n_{\ell m}
\ket{n_{\ell m}(\omega)}
\ ,
\label{Ln}
\ee
where $\ket{n_{\ell m}(\omega)}=(n!)^{-1/2}\left[\hat a^\dagger_{\ell m}(\omega)\right]^n\ket{0}$
(with $n=n_{\ell m}$ for brevity).
\par
Let us consider a coherent state of fixed $\omega$ (which we omit for simplicity), $\ell$ and $m$,
\be
\ket{g_{\ell m}}
&\!\!=\!\!&
e^{-N_{\ell m}/2}\,
\exp\left\{
g_{\ell m}\, \hat a^{\dagger}_{\ell m}
\right\}
\ket{0}
\nonumber
\\
&\!\!=\!\!&
e^{-N_{\ell m}/2}\,
\sum_n
\frac{\left(g_{\ell m}\,\hat a^\dagger_{\ell m}\right)^n}{n!}
\ket{0}
\nonumber
\\
&\!\!=\!\!&
e^{-N_{\ell m}/2}\,
\sum_n
\frac{g_{\ell m}^n}{\sqrt{n!}}
\ket{n_{\ell m}}
\ .
\label{g_lm}
\ee
From $\pro{n_{\ell m}}{n'_{\ell m}}=\delta_{n n'}$, the normalisation 
\be
\pro{g_{\ell m}}{g_{\ell m}}
=
e^{-N_{\ell m}}\,
\sum_n
\frac{g^{2\,n}_{\ell m}}{n!}
=
1
\ee
implies $N_{\ell m}=g_{\ell m}^2$.
From Eq.~\eqref{Ln}, we then find
\be
\bra{g_{\ell m}}
\sqrt{\hat L^2}
\ket{g_{\ell m}}
&\!\!=\!\!&
e^{-g^2_{\ell m}}
\sum_{n,s}
\frac{g^{s}_{\ell m}}{\sqrt{s!}}\,
\frac{g^{n}_{\ell m}}{\sqrt{n!}}
\bra{s_{\ell m}}
\sqrt{\hat L^2}
\ket{n_{\ell m}}
\nonumber
\\
&\!\!=\!\!&
e^{-g^2_{\ell m}}\,
\sum_{n_{\ell m}}
\frac{g^{2\,n_{\ell m}}_{\ell m}}{n_{\ell m}!}
\hbar\,\sqrt{\ell\,(\ell+1)}\,n_{\ell m}
\nonumber
\\
&\!\!=\!\!&
e^{-g^2_{\ell m}}\,
\hbar\,\sqrt{\ell\,(\ell+1)}
\sum_{n_{\ell m}}
\frac{g^{2\,n_{\ell m}}_{\ell m}}{(n_{\ell m}-1)!}
\nonumber
\\
&\!\!=\!\!&
\hbar\,\sqrt{\ell\,(\ell+1)}\,g^2_{\ell m}\,
e^{-g^2_{\ell m}}\,
\sum_{n}
\frac{g^{2\,n}_{\ell m}}{n!}
\nonumber
\\
&\!\!=\!\!&
\hbar\,\sqrt{\ell\,(\ell+1)}\,N_{\ell m}
\ ,
\ee
which is Eq.~\eqref{expansion of field J^2} with $N_{\ell m}=g^2_{\ell m}(\omega)$.
Likewise,
\be
\bra{g_{\ell m}}
\hat L_z
\ket{g_{\ell m}}
=
\hbar\,m\,N_{\ell m}
\ee
which is Eq.~\eqref{expansion of field J^z}.
\end{document}